\documentclass[conference]{IEEEtran}
\usepackage{caption}
\usepackage{subcaption} 
\usepackage{cite}
\usepackage{url}
\usepackage[breaklinks,colorlinks=true, allcolors=blue]{hyperref}
\usepackage{amsmath}    
\usepackage{amssymb}
\usepackage[linesnumbered,ruled,vlined]{algorithm2e}
\usepackage{graphicx}
\usepackage{textcomp}
\usepackage{lipsum} 
\usepackage{xcolor}
\usepackage{float}
\usepackage{booktabs} 
\usepackage{mdframed}
\usepackage{multicol} 
\usepackage{subcaption}
\usepackage{stfloats}

\usepackage{listings}
\usepackage{xcolor}

\def\BibTeX{{\rm B\kern-.05em{\sc i\kern-.025em b}\kern-.08em
    T\kern-.1667em\lower.7ex\hbox{E}\kern-.125emX}}

\makeatletter 
\newcommand{\linebreakand}{
  \end{@IEEEauthorhalign} 
  \hfill\mbox{}\vspace{3mm}\par
  \mbox{}\hfill\begin{@IEEEauthorhalign}
}
\makeatother 

\begin{document}

\title{Distributional Drift Detection in Medical Imaging with Sketching and Fine-Tuned Transformer\thanks{The authors gratefully acknowledge the support of Dr. Ravi Samala from the U.S. Food and Drug Administration.}}

\author{\IEEEauthorblockN{Yusen Wu$^{\dagger}$, Phuong Nguyen$^\ddagger$, Rose Yesha$^\S$, Yelena Yesha$^\ddagger$}
\IEEEauthorblockA{$^\dagger$Department of Neurology, University of Miami, FL USA\\
$^\ddagger$Department of Computer Science \& IDSC, University of Miami, FL USA\\ 
$^\S$Georgetown University Medical Center, Washington D.C. USA\\ 
\{yxw1259, pnx208, yxy806\}@miami.edu, rose.yesha@medstar.net}}

\IEEEoverridecommandlockouts

\makeatletter
\def\ps@IEEEtitlepagestyle{%
  \def\@oddhead{\parbox[t]{\textwidth}{\centering
      {\footnotesize\color{black}Accepted by IEEE International Conference on Digital Health--- Part of the 2025 IEEE World Congress on SERVICES, Helsinki, Finland}\vspace{1em}}%
  }%
  \def\@oddfoot{}%
}
\makeatother

\maketitle
\IEEEpubidadjcol

\begin{abstract}

Distributional drift detection is important in medical applications as it helps ensure the accuracy and reliability of models by identifying changes in the underlying data distribution that could affect the prediction results of machine learning models. However, current methods have limitations in detecting drift, for example, the inclusion of abnormal datasets can lead to unfair comparisons. This paper presents an accurate and sensitive approach to detect distributional drift in CT-scan medical images by leveraging data-sketching and fine-tuning techniques. We developed a robust baseline library model for real-time anomaly detection, allowing for efficient comparison of incoming images and identification of anomalies. Additionally, we fine-tuned a pre-trained Vision Transformer model to extract relevant features, using mammography as a case study, significantly enhancing model accuracy to 99.11\%. Combining with data-sketches and fine-tuning, our feature extraction evaluation demonstrated that cosine similarity scores between similar datasets provide greater improvements, from around 50\% increased to 99.1\%. Finally, the sensitivity evaluation shows that our solutions are highly sensitive to even 1\% salt-and-pepper and speckle noise, and it is not sensitive to lighting noise (e.g., lighting conditions have no impact on data drift). The proposed methods offer a scalable and reliable solution for maintaining the accuracy of diagnostic models in dynamic clinical environments.

\end{abstract}


\section{Introduction}

Distributional drift \cite{cutler2023stochastic}, also known as dataset drift, in medical imaging refers to changes in data distribution over time, which can significantly affect the accuracy of machine learning models used for diagnostic purposes. This drift may result from various factors, including alterations in imaging equipment, differences in imaging protocols, variations in patient demographics, or updates in image preprocessing techniques. Detecting and managing drift is critical in the medical field to ensure that models remain accurate and reliable. Ignoring drift can lead to incorrect diagnoses or suboptimal treatment recommendations, thereby potentially compromising patient care. Therefore, continuous monitoring and adaptation of models are essential to maintain their effectiveness in dynamic clinical environments. Despite significant advancements in data drift detection within medical imaging, current methodologies exhibit several limitations that must be addressed to improve their effectiveness. We discuss the limitations of \textit{L1}, \textit{L2} and \textit{L3} methods as follows:

(\textit{L1}): Analyzing the distribution of image data is particularly challenging due to its high-dimensional characteristics. Unlike numerical data, where statistical properties like mean and variance can be straightforwardly computed and compared, image data needs specialized feature extraction and dimensional reduction techniques to perform meaningful distribution analysis.

 \begin{figure}[t]
    \centering
    \includegraphics[width=0.45\textwidth]{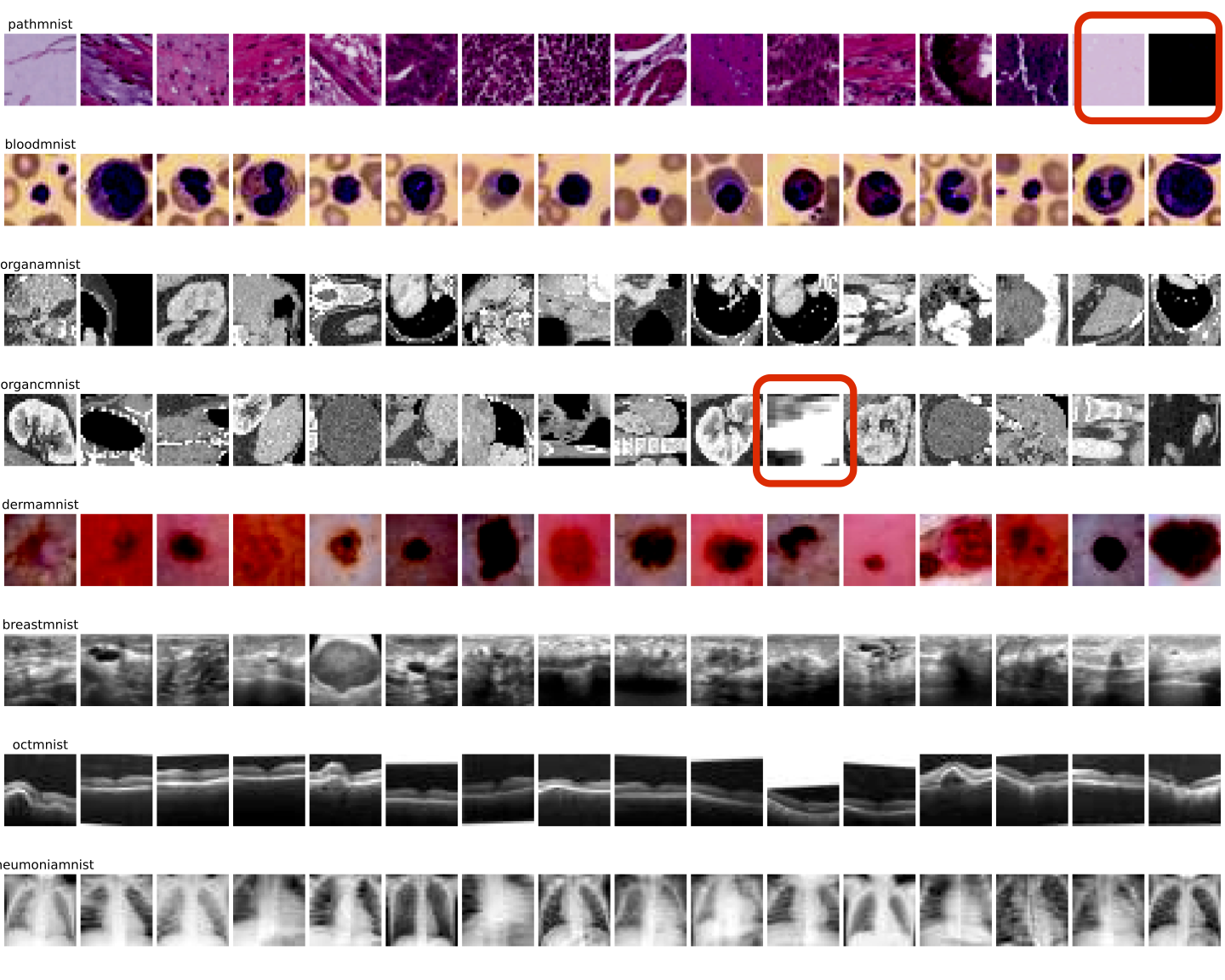}
    \caption{Med-MNIST Benchmarks. We highlight three unqualified images in the benchmarks.}
    \label{fig:abnomal}
\end{figure}

(\textit{L2}): Abnormal datasets can introduce bias in drift detection, which can often be mitigated without compromising the overall data quality. However, efficiently identifying and isolating these abnormal datasets is essential to preserve data integrity. Such abnormalities may arise from data entry errors, sensor malfunctions, or even malicious activities. Implementing effective and efficient anomaly detection techniques is critical for pre-processing data before applying drift detection methods. For instance, as shown in
Fig. \ref{fig:abnomal}, three medical CT images were identified as unqualified for model training.

(\textit{L3}): The absence of real-time processing solutions to handle abnormal data, especially during the pre-processing stage. This process often relies on hospital experts to manually label and remove invalid data before training, which significantly increases the cost and complexity of model development. A solution that enables fast detection is needed.

To overcome the identified limitations, we fine-tuned a pre-trained model specifically optimized for feature extraction from the target images. We fine-tuned Vision Transformer (ViT)\cite{liu2021swin} model with 9,987 mammography images as a case study. The selection of ViT was based on its superior accuracy post-fine-tuning, as demonstrated by our experimental results in the evaluation section. The fine-tuning process of ViT involved several crucial steps. Initially, we loaded the model with weights pre-trained on ImageNet \cite{deng2009imagenet}, a large-scale image dataset, providing a robust foundation for feature extraction. We then replaced the fully connected layers with a custom architecture tailored to our specific task, enabling the model to learn the intricate details of the breast cancer dataset. Additionally, we selectively unfroze the later layers of the model, allowing them to be retrained on our dataset while keeping the earlier layers fixed. This strategy enabled us to fine-tune the model effectively, preserving the general features learned from the larger dataset while adapting it to the unique characteristics of the breast cancer images.

Additionally, our study designed an advanced  \textit{data sketching} \cite{leng2015online,ting2020data} solution, specifically employing MinHash \cite{christiani2017set} as the sketching method, to build a robust baseline library for anomaly detection in image pre-processing phase. This baseline model acts as a reference, allowing systematic comparison of incoming images to identify and correct anomalies within the dataset. By leveraging data sketches, the image data is condensed into a smaller, fixed-size structure, the MinHash signature, which can be quickly compared with other sketches to estimate similarity. A low similarity score may indicate abnormal or low-quality images that require further review.

Finally, after excluding all the low score images, all the extracted features from selected images are rigorously compared with the baseline library to detect drift, using both Kolmogorov-Smirnov \cite{fasano1987multidimensional} and cosine similarity \cite{xia2015learning} algorithms. Our solutions compared with 3 baseline models. The results show that our solution is efficient in identifying subtle changes and deviations within the medical images. 

Our research has yielded several important \textbf{findings}.  
\begin{itemize}
    \item We discovered that data sketches significantly enhance data quality by creating compact, noise-reduced representations, which lead to more stable similarity comparisons. 
    \item We discovered that fine-tuning a ViT pre-trained model proved highly effective, as it greatly improved the quality of feature extraction and boosted similarity scores, resulting in more fair and relevant comparisons.
    \item Our sensitivity evaluation revealed that the combination of data sketches and a fine-tuned pre-trained model is capable of efficiently detecting distributional drift, even when the data contains as little as 1\% Salt-and-Pepper and Speckle random noise (transmission errors, intentional tampering, subject variations, etc.). Moreover, lighting conditions have little to no impact on data drift.
    \item Our solution is applicable to a wide range of medical imaging scenarios that require prompt drift detection.
\end{itemize}

The reminder of this paper is organized as follows: Section \ref{background} provides an overview of the relevant background. Section \ref{pm} details our proposed methods. Section \ref{er} presents the experimental results, while Section \ref{rw} discusses related work. Finally, we conclude our research in Section \ref{con}.

\section{Background} \label{background}

\subsection{Distributional Drift (Dataset Shift) in Medical Images}
 Distributional drift, also known as dataset shift, refers to changes in the statistical properties of data between the training and deployment phases of machine learning models \cite{cutler2023stochastic}. In medical imaging, this drift can manifest as differences in pixel intensities, textures, or other visual characteristics between the images used for training and those encountered during real-world application. This phenomenon is particularly prevalent in medical imaging due to the continuous evolution of imaging technologies, variations in imaging equipment, and changes in patient demographics. Such variations can result in significant differences between the training data and new data, challenging the model’s ability to make accurate predictions.

Addressing distributional drift is critical because it directly affects the performance and reliability of machine learning models in clinical settings. A model trained on a specific dataset may perform well during initial testing but might fail to maintain its accuracy when applied to new datasets that differ in subtle yet significant ways. This discrepancy can lead to misdiagnosis or incorrect treatment recommendations, potentially compromising patient safety. As medical imaging practices and technologies continue to evolve, it is essential to continuously monitor and adapt models to ensure their effectiveness. Ensuring that models can robustly handle or adapt to distributional drift is vital not only for maintaining high standards of patient care but also for complying with regulatory requirements governing the use of AI in healthcare.

\subsection{Causes of Distributional Drift in Image Data} \label{factors}
Distributional drift in image data can occur due to various factors related to changes in data, environmental conditions, or the image acquisition process. One common cause is variations in lighting conditions, which can occur due to different times of day, weather conditions, or light sources, leading to alterations in brightness and contrast that affect the data distribution. Changes in imaging devices, such as switching cameras or scanning devices, inconsistent calibration, lens aging, or fluctuations in sensor performance, can also result in changes to image quality and subsequently impact the data distribution. Another contributing factor is subject variations, where changes in the shape, posture, or size of the subject being imaged can lead to shifts in the feature distribution of the images. Additionally, modifying the resolution or size of images can alter the distribution of image features, potentially causing drift. Different devices or sensors capturing the same scene may record varying levels of detail, leading to distributional drift. Finally, images taken from different angles of the same subject can introduce changes in viewpoint, resulting in shifts in the data distribution. These factors highlight the complexity of managing distributional drift in image data and the need for robust strategies to mitigate its impact on model performance.

\section{Related Work} \label{rw}
Existing research on distributional drift detection has predominantly focused on textual data, leveraging changes in word embeddings \cite{diab2023natural} or similar features to monitor drift \cite{sahiner2023data, yu2022meta, madaan2023detail}. While these methods have demonstrated effectiveness in text-based applications, they face significant challenges when applied to high-dimensional medical imaging data. Unlike text, subtle variations in pixel intensities or textures in images can have substantial clinical implications, demanding more sensitive and robust approaches.

Several studies have explored transfer learning to adapt pre-trained models to new datasets \cite{feldhans2021drift,bjerva2020back,li2022drift,grundmann2024data}. Transfer learning effectively reduces the need for large training datasets, but it is prone to errors from noisy or outlier data. The reliance on models that are highly sensitive to such imperfections limits their capacity to detect minor distributional changes, which is crucial in medical applications. Furthermore, these approaches often overlook real-time requirements, making them unsuitable for dynamic clinical environments where prompt drift detection is critical.

Drift detection in medical imaging remains underexplored compared to text data. Existing methods typically involve visual inspection or traditional image processing techniques \cite{mera2019incremental,rahmani2023assessing,yeshchenko2020vdd}, which are often not scalable for large datasets or capable of handling the complexity of modern imaging data. For instance, approaches relying solely on statistical tests or basic similarity measures fail to capture deeper relationships within high-dimensional image datasets, leading to inconsistent and unreliable drift detection results.

Our proposed method addresses these limitations by integrating fine-tuning of a ViT model with advanced data sketching techniques. This combination enhances the robustness and sensitivity of drift detection, even in the presence of minimal noise. Unlike methods that rely heavily on statistical metrics alone, our approach leverages the strengths of data sketches to efficiently summarize high-dimensional data and employs cosine similarity to identify subtle distributional shifts. Additionally, by incorporating real-time anomaly detection mechanisms, our method is well-suited for dynamic clinical settings, offering a scalable and reliable solution for maintaining diagnostic model performance.

In contrast to prior approaches, which often struggle with noisy or imbalanced datasets, our method demonstrates resilience by filtering low-quality data and optimizing feature extraction. By achieving stable similarity comparisons and high sensitivity to minor anomalies, our solution ensures more accurate drift detection, addressing key gaps in existing methodologies.

\section{Proposed Methods} \label{pm}

\begin{figure*}[t]
    \centering
    \includegraphics[width=0.8\textwidth]{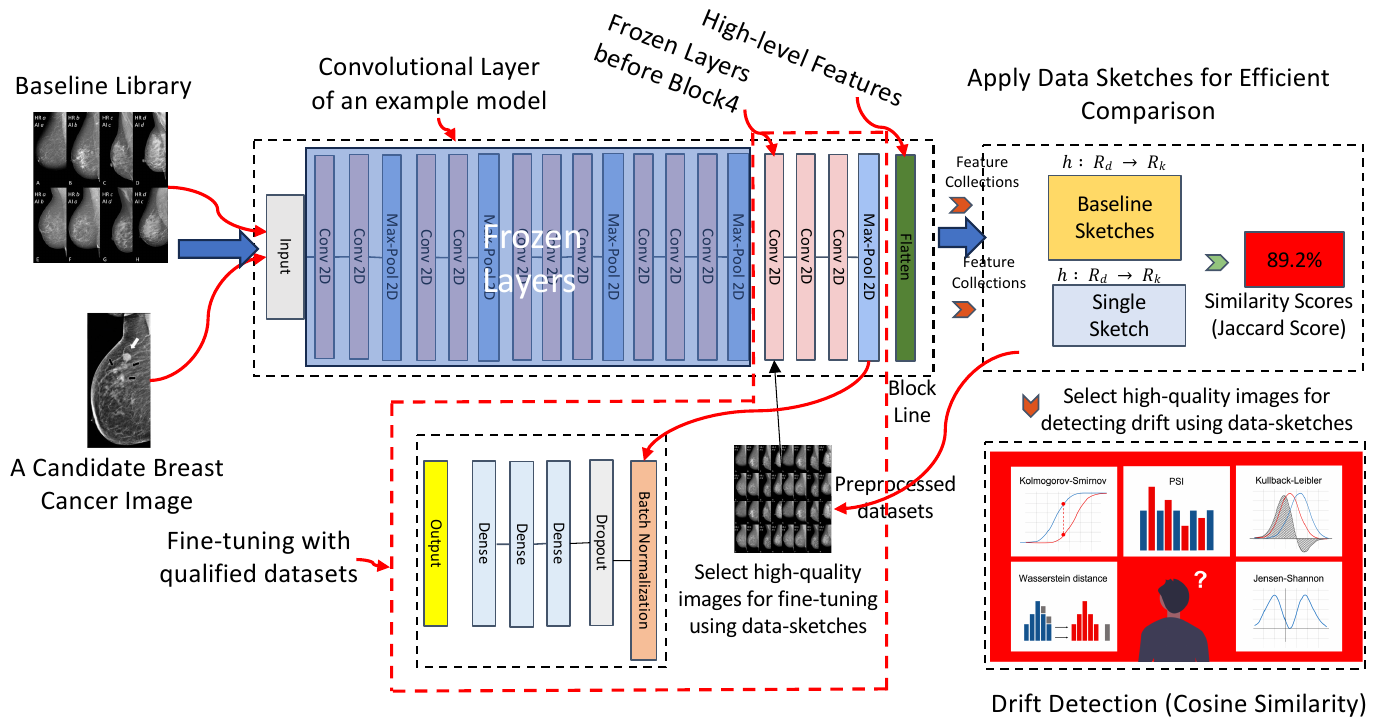}
    \caption{Workflow of our data-sketches-based and fine-tuned pre-train model for drift detection.}
    \label{fig:report0}
\end{figure*}

To effectively address the challenges of distributional drift in medical imaging and enhance the accuracy of feature extraction, we propose a comprehensive solution that combines advanced fine-tuning techniques with real-time anomaly detection using data sketching, as shown in Fig. \ref{fig:report0}. This approach not only optimizes model performance but also ensures the reliable detection of subtle changes in data distribution within dynamic clinical environments.

\subsection{Fine-tuned Pre-trained Model for Feature Extractions}

Our method leverages a pre-trained deep learning model for feature extraction, capitalizing on its ability to encode rich feature representations from input data. Pre-trained models, trained on extensive datasets, offer the advantage of transfer learning, allowing the knowledge gained from large-scale data to be applied to specific tasks with limited data availability. This approach utilizes the pre-trained model within the context of our problem domain to enhance feature extraction.

\subsubsection{Data Preparation}

The first step in fine-tuning a model for medical image drift detection involves thorough data preparation. This process includes selecting an qualified dataset, preprocessing the images, and applying data-sketches to improve the data quality.

\subsubsection{Layer Freezing and Fine-Tuning}

To preserve the general features learned during pretraining, the some of the layers from the pre-trained model are frozen, not updated during the fine tuning process. Only the final few layers, which are responsible for task-specific features, are fine-tuned. Let $\mathbf{W} = \{\mathbf{w}_1, \mathbf{w}_2, \ldots, \mathbf{w}_n\}$ be the pre-trained model parameters, and $n \in \mathbb{Z}^+$ is the index of the layers with trainable parameters. Supposing the data flow is from the layer-$1$ propagating till layer-$n$, we can have the first $l$-frozen layers and regular layers as
\begin{align}
   \mathbf{W}_{\text{frozen}} &= \{ \mathbf{w}_1,  \mathbf{w}_2, \ldots,  \mathbf{w}_l\} \\
   \mathbf{W}_{\text{tunable}} &= \{ \mathbf{w}_{l+1},  \mathbf{w}_{l+2}, \ldots,  \mathbf{w}_n\}
\end{align}
where $\mathbf{W}_{\text{frozen}}$ are the weights of the frozen layers, and $\mathbf{W}_{\text{tunable}}$ are the weights of the layers to be fine-tuned. The pre-trained models \cite{dosovitskiy2021imageworth16x16words,
he2015deepresiduallearningimage,
simonyan2015deepconvolutionalnetworkslargescale} are targeting on other tasks rather than detecting the data drift. Therefore, the minor customization are added to the final layer. It is replaced with a fully connected layer with sigmoid function as the activation function, tailored for the binary classification task:
\[
p = \frac{1}{1+e^{-\mathbf{\bar w}^{\top}_n\mathbf{x}_n + \bar b}}
\]
where $\mathbf{x}_n$ corresponds to output embedding from the previous layer, ${\top}$ denotes the vector transpose and $\mathbf{\bar w}_n$ and $\bar b \in \mathbb{R}$ consist of $\mathbf{w}_n$, the parameters of the last layer.

\subsubsection{Loss Function}

The loss function used for training is the Binary Cross-Entropy (BCE) loss, which is suitable for binary classification tasks. Given the output probability $p$ from the final layer, the BCE loss is defined as:
\[
\mathcal{L}_{\text{BCE}} = -\frac{1}{B} \sum_{i=1}^{B} [y_i \log(p_i) + (1-y_i) \log(1-p_i)]
\]
where $y_i$ is the true label (1 for the correct class, 0 otherwise), $p_i$ is the predicted probability for training sample $i$ and $B \in \mathbb{Z}^+$ is the size of the input, typically as the mini-batch size.

\subsubsection{Optimization Algorithm}

We use mini-batch stochastic gradient descent employing the Adam optimizer for training, which adjusts the learning rate for each parameter individually based on estimates of lower-order moments. The update rule for the parameters is given by:
\[
\mathbf{W}_t = \mathbf{W}_{t-1} - \eta \cdot \frac{\mathbf{m}_t}{\sqrt{\mathbf{v}_t} + \mathbf{\epsilon}}
\]
where $\mathbf{W}_t$ are the model parameters at step $t$, $\eta$ is the learning rate, $\mathbf{m}_t$ is the first moment estimate (mean of gradients), and $\mathbf{v}_t$ is the second moment estimate (uncentered variance of gradients). $\epsilon$ is a small constant to prevent division by zero.

\subsubsection{Training Setup}

The model is fine-tuned over a series of epochs, and we set it as 20 as the learning curve show the convergence, which is also depending on the dataset size and complexity. The training involves iterating over the dataset in mini-batches, computing the loss for each batch, and updating the model parameters.

The learning rate is initially set to a low value $5 \times 10^{-5}$, as we chosen in the experiments,  to avoid drastic changes to the pre-trained weights. The learning rate may be reduced further if the validation performance plateaus.

\subsection{Data Sketches for Real-time Anomaly Detection}

Data sketches are efficient probabilistic data structures that provide approximate representations of large datasets. These structures allow for scalable computations of various statistical properties without requiring the entire dataset to be stored or processed at once. In this subsection, we will delve into the mathematical foundation of data sketches and explain our anomaly detection steps. We summarized our method in Algorithm \ref{alg:algrthm}.

\subsubsection{Feature Extraction and Hashing}

Given a set of images $\mathcal{I} = \{I_1, I_2, \ldots, I_m\}$, we first extract feature vectors $\mathbf{f}_i \in \mathbb{R}^d$ from each image $I_i$ using a pre-trained model. Let $\mathcal{F} = \{\mathbf{f}_1, \mathbf{f}_2, \ldots, \mathbf{f}_m\}$ represent the set of extracted features and the feature extractor, namely the corresponding layers of the neural network, are noted as $\mathbf{\phi}$
\[
\mathbf{f}_i = \mathbf{\phi}(I_i), \quad \forall I_i \in \mathcal{I}
\]
To create a data sketch, we apply a hash function $h: \mathbb{R}^d \rightarrow \mathbb{R}^k$ that maps each feature vector to a lower-dimensional space.
\[
\mathbf{h}_i = h(\mathbf{f}_i), \quad \mathbf{h}_i \in \mathbb{R}^k, \quad \forall \mathbf{f}_i \in \mathcal{F}
\]
Hash functions are well-performed in clustering and anomaly detection, because they can be used to group similar data points together and detect anomalies in data by identifying data points with unusual hash values.

\subsubsection{Constructing the Baseline Library}

The baseline library $\mathcal{B}$ is constructed by aggregating the hashed feature vectors of a representative set of CT-scan images. Let $\mathcal{F}_\text{baseline}$ denote the set of features from the baseline images.
\[
\mathcal{B} = \{ \mathbf{h}_1, \mathbf{h}_2, \ldots, \mathbf{h}_m \}, \quad \mathbf{h}_i = h(\mathbf{f}_i), \quad \forall \mathbf{f}_i \in \mathcal{F}_\text{baseline}
\]

\subsubsection{Integration with Real-time Feature Extraction}

Our approach seamlessly integrates data sketches with feature extraction to enhance the accuracy and efficiency of drift detection. The pre-trained model extracts intricate features from each image, converting them into data sketches that allow for quick and scalable comparisons. This integration ensures that our method can handle high-dimensional medical images while maintaining high performance in real-time applications.

By leveraging the power of data sketches and the KS statistic, we provide a comprehensive solution for detecting and managing data drift in CT-scan medical images, ensuring the reliability and accuracy of diagnostic models.

\subsubsection{Real-time Anomaly Detection}

For an incoming image $I_\text{new}$, we extract its feature vector $\mathbf{f}_\text{new}$ and hash it to obtain the corresponding data sketch $\mathbf{h}_\text{new}$.
\[
\mathbf{f}_\text{new} = \mathbf{\phi}(I_\text{new})
\]
\[
\mathbf{h}_\text{new} = h(\mathbf{f}_\text{new})
\]
To detect data drift, we compare the distribution of the new data sketch $\mathbf{h}_\text{new}$ with the baseline library $\mathcal{B}$ using the Jaccard similarity. The Jaccard score measures the maximum difference between the empirical cumulative distribution functions (CDFs) of the two datasets.

\begin{algorithm}[t] 
\caption{Real-time Detection of Abnormal or Poor-Quality Datasets using Data Sketches}
\label{alg:algrthm}
\KwIn{Baseline library $\mathcal{B}$, incoming image $I_{\text{new}}$}
\KwOut{Quality detection result}
Extract feature vector $\mathbf{f}_{\text{new}}$ from $I_{\text{new}}$\;
Hash the feature vector to obtain data sketch $\mathbf{h}_{\text{new}}$\;
Compute the Jaccard similarity $J(\mathbf{h}_{\mathcal{B}}, \mathbf{h}_{\text{new}})$ between the baseline library $\mathcal{B}$ and the new image $I_{\text{new}}$\;
\eIf{$J(\mathbf{h}_{\mathcal{B}}, \mathbf{h}_{\text{new}}) < J_{\alpha}$}{
    \KwRet{Abnormal or poor-quality data detected}\;
}{
    \KwRet{Data quality acceptable}\;
}
\end{algorithm}

\subsection{Similarity Comparisons for Drift Detection}

\subsubsection{Kolmogorov-Smirnov Statistic} \label{kspvalue}
The Kolmogorov-Smirnov (KS) test is a non-parametric statistical method used to compare the cumulative distribution functions (CDFs) of two datasets or a dataset and a theoretical distribution. It evaluates the maximum distance between the distributions (D-statistic) and determines whether the observed difference is statistically significant. The p-value associated with the KS test indicates the likelihood of observing the difference under the null hypothesis, which assumes the datasets come from the same distribution. A low p-value (e.g., $<$ 0.05) suggests a significant difference, while a high p-value indicates no evidence to reject the null hypothesis, implying similarity between distributions.

Formal definition: Let $F_\mathcal{B}(x)$ and $F_\text{new}(x)$ be the empirical CDFs of the baseline library and the new data sketch, respectively. The KS statistic $D$ is defined as:
\[
D = \sup_x | F_\mathcal{B}(x) - F_\text{new}(x) |
\]
where $\sup$ denotes the supremum function. The KS test then evaluates whether $D$ exceeds a critical value $D_\alpha$ based on the desired significance level $\alpha$.

\subsubsection{Cosine Similarity} \label{cosine}

Cosine similarity is a useful metric for comparing two medical images by evaluating the similarity between their feature vectors. Each image is represented as a vector of features, and the cosine similarity between these vectors is calculated as:
\[
\text{cosine\_similarity}(\mathbf{a}, \mathbf{b}) = \frac{\mathbf{a}^{\top}\mathbf{b}}{\|\mathbf{a}\| \|\mathbf{b}\|}
\]
Here, \(\mathbf{a} \in \mathbb{R}^{M}\) and \(\mathbf{b} \in \mathbb{R}^{M}\) are the feature vectors of the two images and $M \in \mathbb{Z}^{+}$ is arbitrary. The value ranges from -1 to 1, with 1 indicating identical images, 0 indicating no similarity, and -1 indicating completely negative identical images.

In the context of medical imaging, including CT scans and MRIs, cosine similarity is valuable for tasks like identifying abnormalities or monitoring disease progression. It quantifies the similarity between a new image and either a reference image or a collection of normal images. This metric is particularly effective in detecting subtle changes in image features, which are crucial for accurate diagnosis and treatment planning.

\subsection{Sensitivity Evaluation via Incremental Noise Introduction}
To assess the sensitivity of cosine similarity in detecting data drift, we systematically introduced incremental noise into our experimental dataset and monitored the similarity metric. Specifically, Gaussian noise was added in 10-20\% increments to simulate real-world disturbances in medical imaging data, such as fluctuations in image acquisition or processing conditions.  Beginning with a dataset free of noise, we gradually added noise at each phase and recalculated the cosine similarity for each pair of images following each increase. This approach allowed us to methodically evaluate how the similarity scores adjusted to escalating levels of data corruption. Our objective was to pinpoint the threshold at which the cosine similarity metric signaled a significant deviation from the baseline, thus indicating the onset of drift.


\section{Experimental Results} \label{er}
\subsection{Experimental Datasets}

In our research, we utilize the MedMNIST \cite{yang2021medmnist} dataset as benchmark and 9,987 mammography images for fine-tuning. The MedMNIST is an extensive repository of standardized biomedical images specifically created to support the development and assessment of machine learning algorithms in medical imaging. MedMNIST mirrors the structure of the MNIST dataset but is significantly larger and specifically adapted for biomedical purposes. It comprises ten distinct 2D datasets and eight 3D datasets. The ten datasets in MedMNIST encompass a range of medical conditions and imaging techniques. For our analysis, we have selected eight common 2D datasets as benchmarks including PathMNIST, BloodMNIST, OrganMNIST, ChestMNIST, DermaMNIST, BreastMNIST, OCTMNIST and PneumoniaMNIST.

 \begin{figure*}[t]
    \centering
    \begin{subfigure}[b]{0.24\textwidth}
        \centering
        \includegraphics[width=\textwidth]{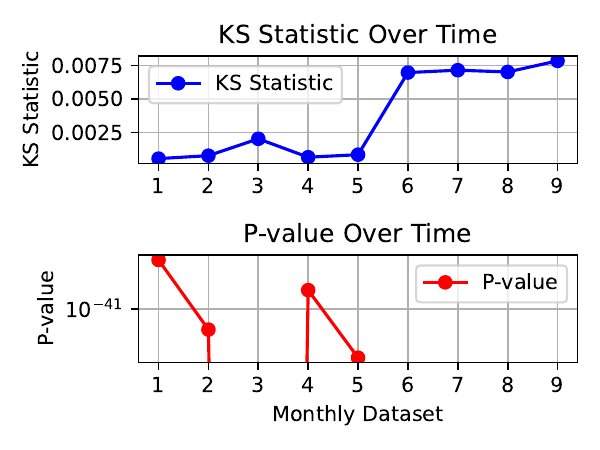}
        \caption{PathMNIST}
        \label{fig:report1}
    \end{subfigure}
    \hfill
    \begin{subfigure}[b]{0.24\textwidth}
        \centering
        \includegraphics[width=\textwidth]{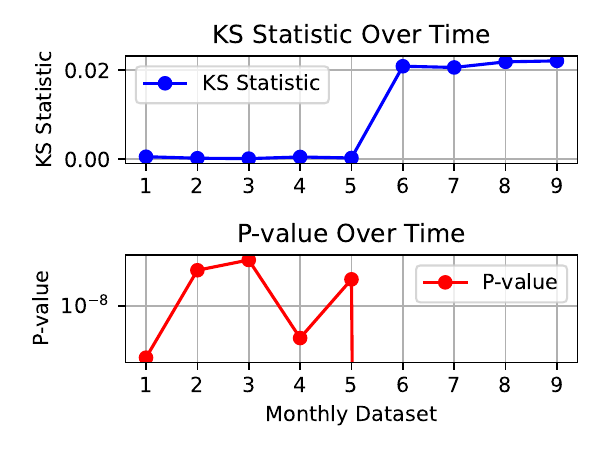}
        \caption{BloodMNIST}
        \label{fig:report2}
    \end{subfigure}
    \hfill
    \begin{subfigure}[b]{0.24\textwidth}
        \centering
        \includegraphics[width=\textwidth]{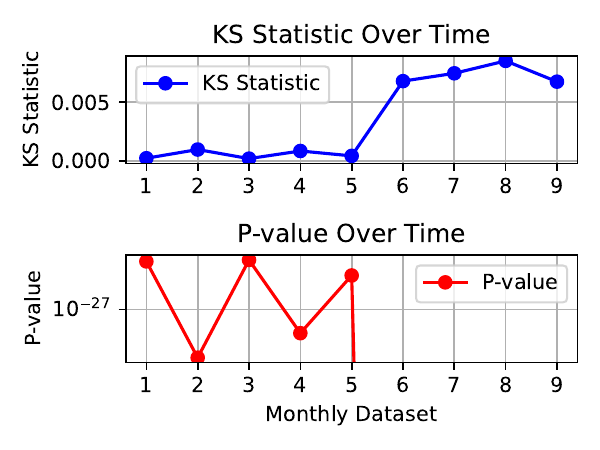}
        \caption{OrganMNIST}
        \label{fig:report3}
    \end{subfigure}
    \hfill
    \begin{subfigure}[b]{0.24\textwidth}
        \centering
\includegraphics[width=\textwidth]{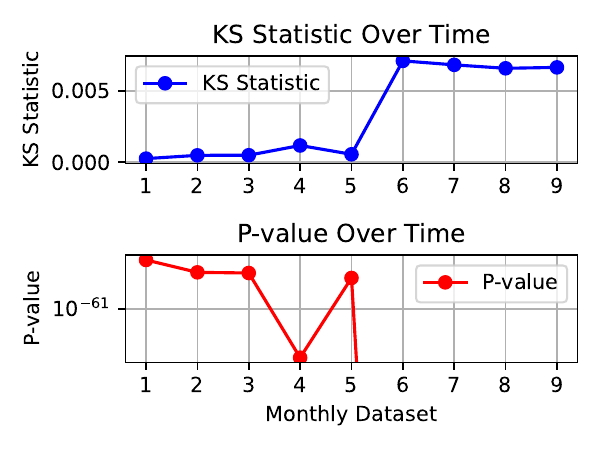}
        \caption{ChestMNIST}
        \label{fig:report4}
    \end{subfigure}

        \begin{subfigure}[b]{0.24\textwidth}
        \centering
        \includegraphics[width=\textwidth]{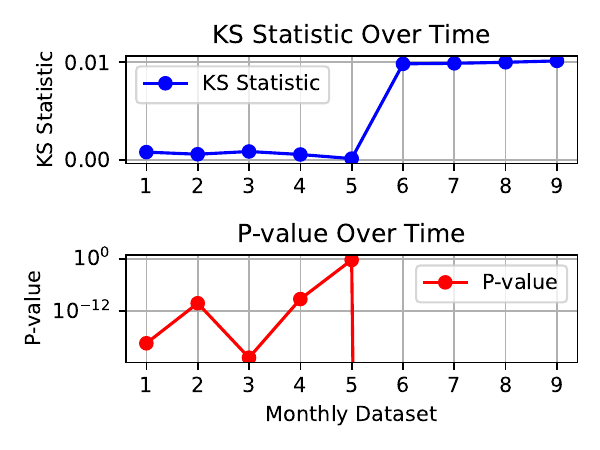}
        \caption{DermaMNIST}
        \label{fig:rep12}
    \end{subfigure}
    \hfill
    \begin{subfigure}[b]{0.24\textwidth}
        \centering
        \includegraphics[width=\textwidth]{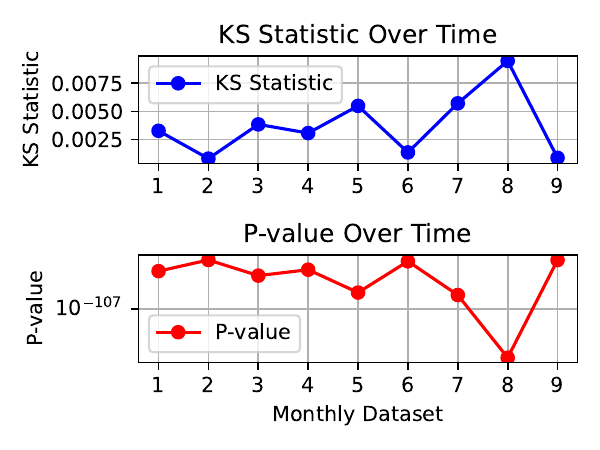}
        \caption{BreastMNIST}
        \label{fig:report2}
    \end{subfigure}
    \hfill
    \begin{subfigure}[b]{0.24\textwidth}
        \centering
        \includegraphics[width=\textwidth]{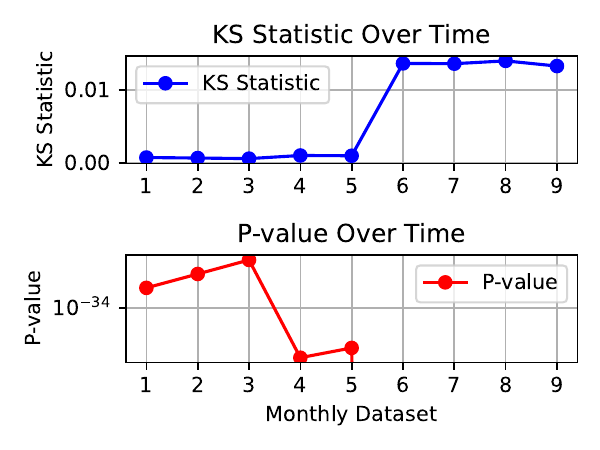}
        \caption{OCTMNIST}
        \label{fig:report3}
    \end{subfigure}
    \hfill
    \begin{subfigure}[b]{0.24\textwidth}
        \centering
        \includegraphics[width=\textwidth]{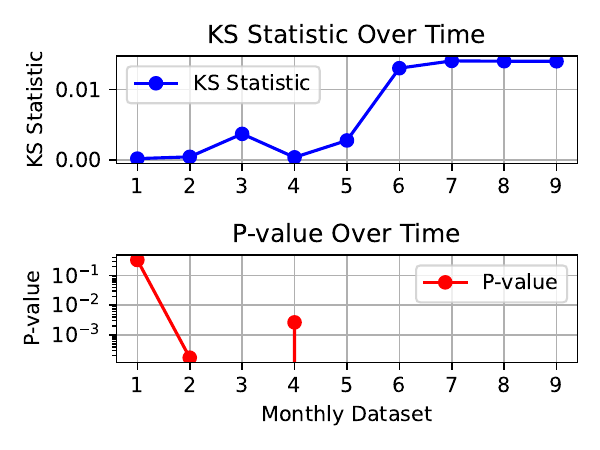}
        \caption{PneumoniaMNIST.}
        \label{fig:report4}
    \end{subfigure}
    \caption{\textbf{Baseline 1 (\textbf{without fine-tuning}):} KS Statistic and P-value (\ref{kspvalue}) Trends Across 8 Med-MNIST Datasets. P-value shows that all the datasets are $<$ 0.05, meaning the low similarities in data distribution. KS statistics trends for the BreastMNIST dataset highlights a critical challenge in identifying data drift within highly variable datasets. Unlike other MNIST variants, the trends in BreastMNIST exhibit significant fluctuations across time, which complicates the interpretation of when drift occurs. The instability in KS statistics suggests that the dataset may be influenced by factors such as inconsistencies in data acquisition processes or intrinsic variability in the imaging data.}
    \label{fig:multi_report}
\end{figure*}

 \begin{figure*}[t]
    \centering
    \begin{subfigure}[b]{0.23\textwidth}
        \centering
        \includegraphics[width=\textwidth]{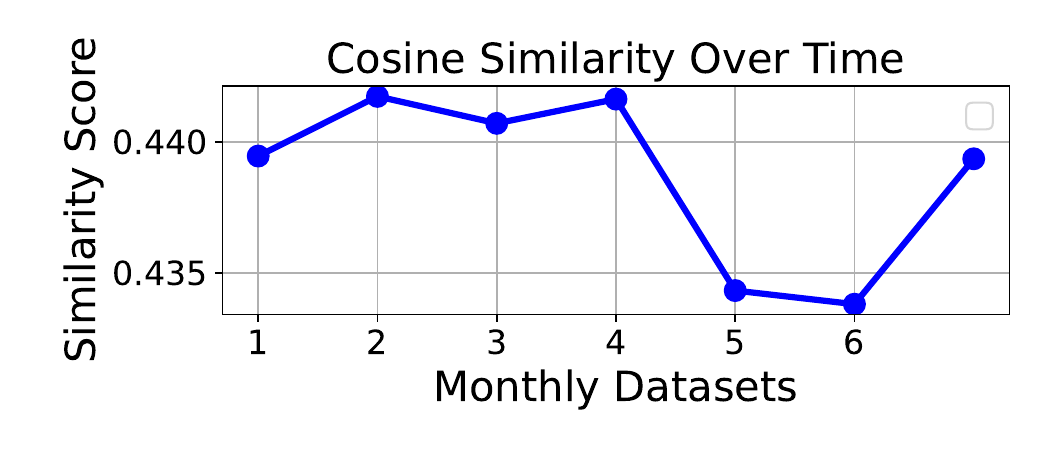}
        \caption{PathMNIST}
        \label{fig:report21}
    \end{subfigure}
    \hfill
    \begin{subfigure}[b]{0.23\textwidth}
        \centering
        \includegraphics[width=\textwidth]{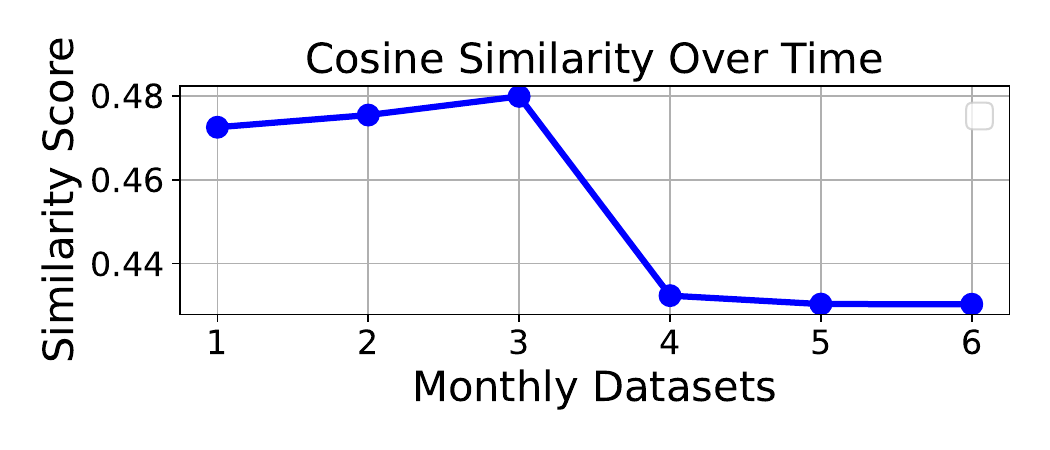}
        \caption{BloodMNIST}
        \label{fig:report2}
    \end{subfigure}
    \hfill
    \begin{subfigure}[b]{0.23\textwidth}
        \centering
        \includegraphics[width=\textwidth]{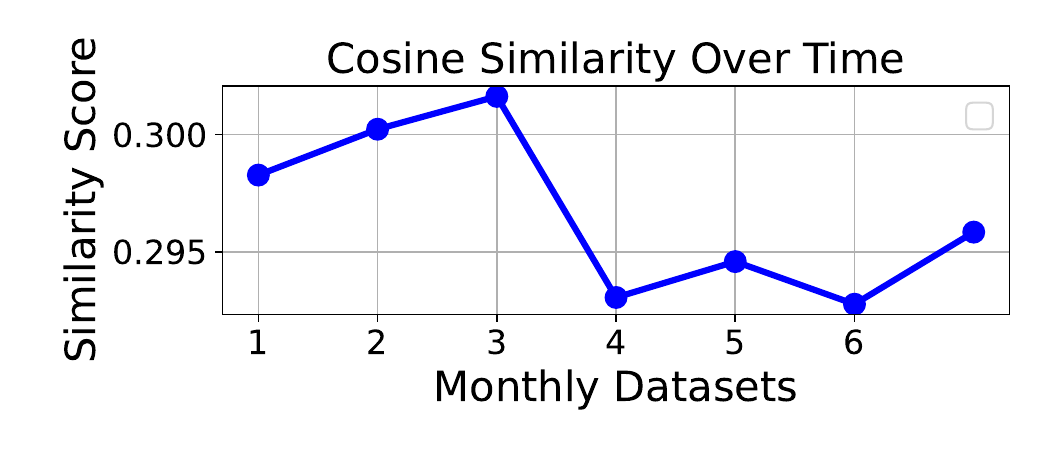}
        \caption{OrganMNIST}
        \label{fig:report3}
    \end{subfigure}
    \hfill
    \begin{subfigure}[b]{0.23\textwidth}
        \centering
        \includegraphics[width=\textwidth]{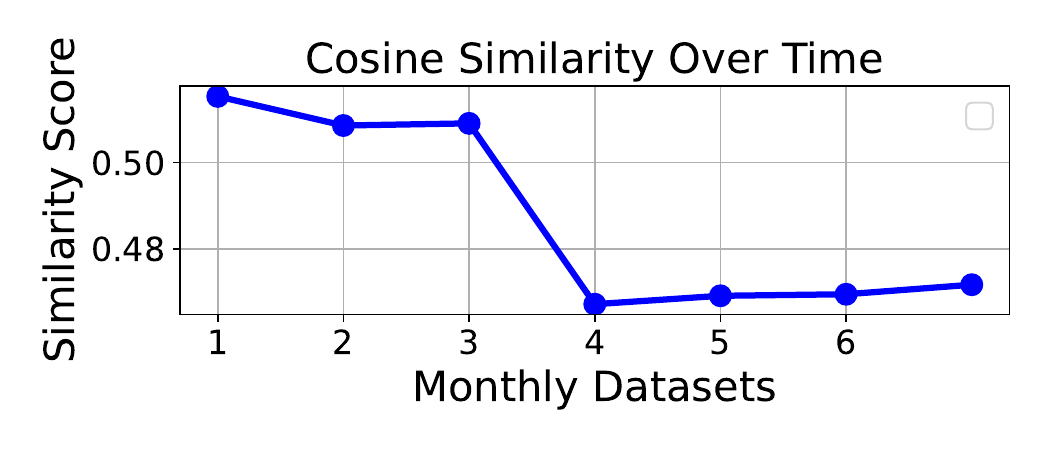}
        \caption{ChestMNIST}
        \label{fig:report4}
    \end{subfigure}

    \begin{subfigure}[b]{0.23\textwidth}
        \centering
        \includegraphics[width=\textwidth]{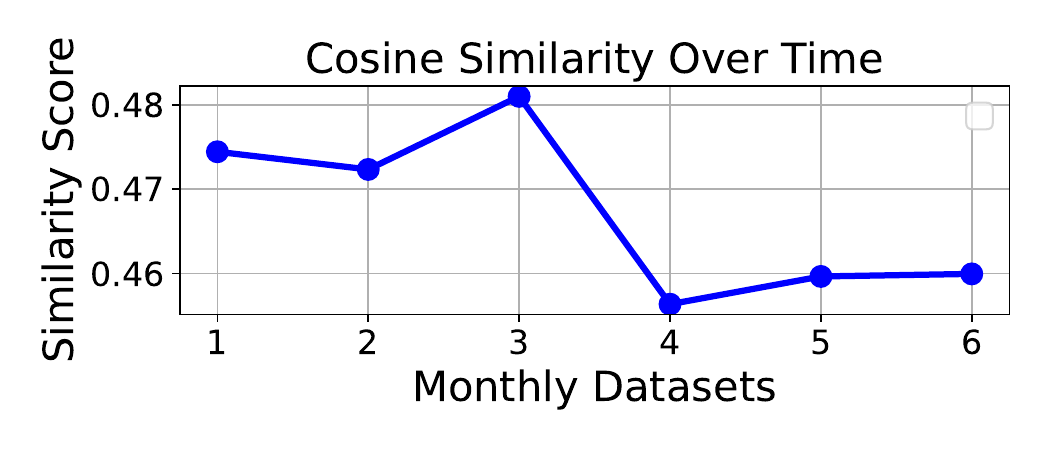}
        \caption{DermaMNIST}
        \label{fig:report78}
    \end{subfigure}
    \hfill
    \begin{subfigure}[b]{0.23\textwidth}
        \centering
        \includegraphics[width=\textwidth]{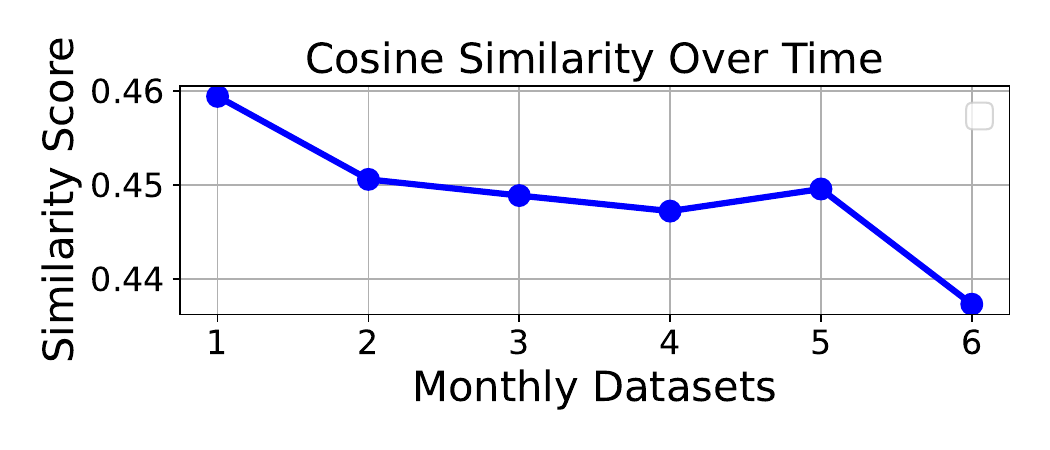}
        \caption{BreastMNIST}
        \label{fig:report2}
    \end{subfigure}
    \hfill
    \begin{subfigure}[b]{0.23\textwidth}
        \centering
        \includegraphics[width=\textwidth]{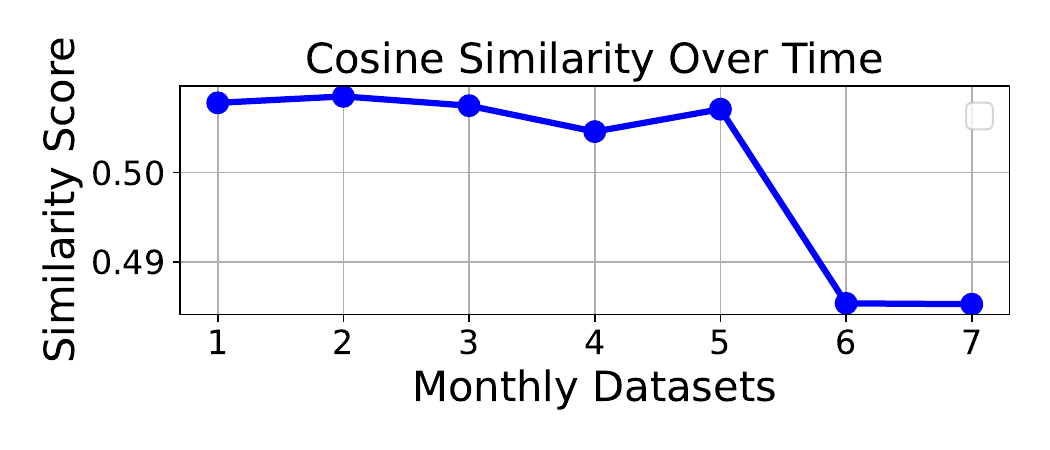}
        \caption{OCTMNIST}
        \label{fig:report3}
    \end{subfigure}
    \hfill
    \begin{subfigure}[b]{0.23\textwidth}
        \centering
        \includegraphics[width=\textwidth]{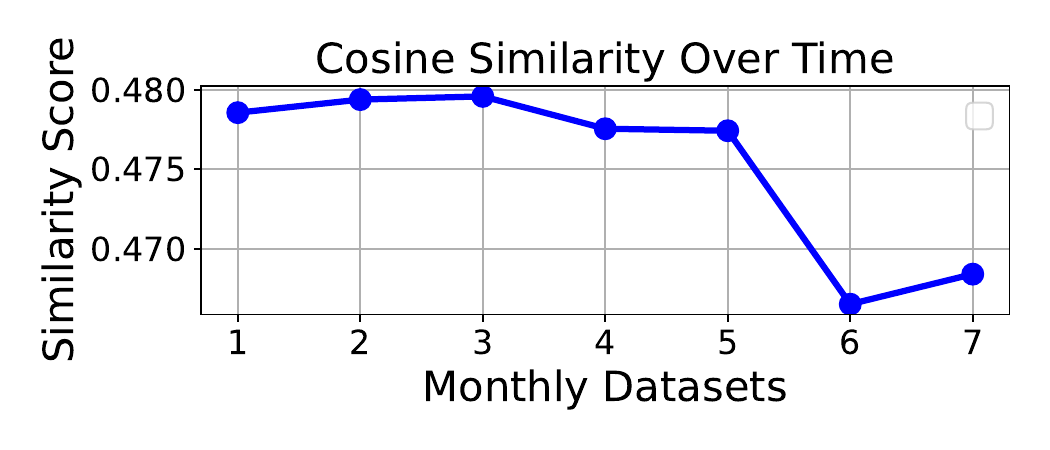}
        \caption{PneumoniaMNIST}
        \label{fig:report4}
    \end{subfigure}
    \caption{\textbf{Baseline 2 (\textbf{without fine-tuning}):} Cosine Similarity Score (\ref{cosine}) Across 8 Med-MNIST Datasets. The analysis of cosine similarity scores across the 8 Med-MNIST datasets reveals a fundamental limitation of feature extraction without fine-tuning. The similarity scores remain consistently low, hovering around 50\%, which indicates that the extracted features fail to capture meaningful relationships between datasets. This low level of similarity suggests that the pre-trained model, when used without fine-tuning, is unable to adapt to the specific characteristics of medical imaging data.}
    \label{fig:multi_report2}
\end{figure*}

\subsubsection{Data Splits}

To ensure robust evaluation, each dataset in MedMNIST is split into 7 to 10 groups for simulating different scenarios of data distributional shifts.

\subsection{Evaluation of Model Accuracy}

\begin{table}[ht]
\centering
\caption{Comparison of Model Test Accuracy}
\begin{tabular}{cc}
\toprule
\textbf{Model} & \textbf{Fine-Tuned Accuracy} \\ \midrule
Customized Model                         & 78.7\%                                                     \\
ResNet50                       & 66.9\%                          \\
ResNet152                           & 66.7\%                          \\
VGG16                          & 92.2\%                                                   \\ 
Vision Transformer(ViT)                          & 99.11\%                                                    \\ 

\bottomrule                        
\end{tabular}
\label{tab:model_accuracy_comparison}
\end{table}

We evaluated the performance of 5 distinct models, Customized Model, VGG16, and ResNet50, ResNet152 \cite{szegedy2017inception} and ViT (patch16)\cite{liu2021swin}, on breast cancer image classification, as shown in Table \ref{tab:model_accuracy_comparison}. The primary goal was to assess the effectiveness of fine-tuning when applied to pre-trained models in comparison to a baseline CNN model trained from scratch.

\textbf{Model Performance.} Each model was fine-tuned on the labeled breast cancer image dataset. The baseline CNN model provided a point of reference for evaluating the improvements gained through more complex architectures like VGG16, ResNet50, and ViT. 

The results clearly demonstrated that fine-tuning yielded superior performance across all models without ResNet50 and ResNet152, with ViT showing particularly strong results up to 99.11\%. Despite fine-tuned ResNet152's deeper architecture, which theoretically should offer better accuracy, its performance did not surpass that of VGG16. 
The performance of ResNet152, despite its depth, can be attributed to the relatively small size of the dataset. Deep architectures like ResNet50 or 152 often require large amounts of data to fully leverage their capacity and avoid overfitting. In contrast, VGG16 and ViT, with their more moderate depth, appeared to be better suited to our dataset's size, allowing it to generalize more effectively during training.

\subsection{Real-time Detection with New Images}

 \begin{figure}[t]
    \centering
    \includegraphics[width=0.28\textwidth]{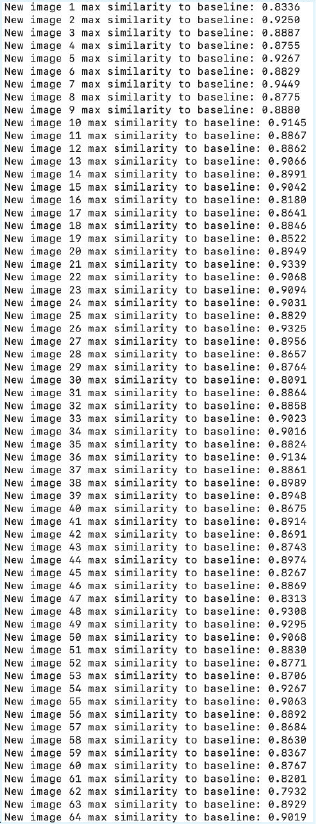}
    \caption{Real-time image similarity comparison}
    \label{fig:report5555555}
\end{figure}

Fig. \ref{fig:report5555555} shows individual images compared with the baseline sketch model. Each image is assigned a similarity score, where a lower score may suggest dissimilar images that require further inspection. Therefore, such images should be highlighted, and the doctor must manually define an appropriate threshold to identify and review them.

\subsection{Evaluation of Drift Detection}
 
\textbf{Baseline 1} (without data-sketches and fine-tuning) Fig. \ref{fig:multi_report} reveals significant trends in the KS statistics and p-values across multiple datasets, providing insights into the onset and progression of data drift. From Fig.~3, it is evident that the KS statistic for BreastMNIST increases steadily over time, starting from approximately 0.003 in the third month and peaking at 0.007 by the sixth month. This upward trend strongly suggests a substantial change in the underlying data distribution.

For PathMNIST and BloodMNIST, a similar rise in the KS statistic is observed, with values exceeding 0.005 around the fourth month. These consistent increases indicate the emergence of distributional drift across these datasets. Importantly, the corresponding p-values for all datasets remain consistently below 0.05, and in some cases, such as BreastMNIST, they fall below 0.01, indicating that the observed differences in data distributions are statistically significant.

The variability in the KS statistic trends across datasets highlights the complex nature of data drift in medical imaging. For BreastMNIST, the KS statistic shows substantial fluctuations month-to-month, which could be attributed to factors such as variations in imaging protocols, patient demographics, or inherent noise in the data collection process. In contrast, datasets like PathMNIST exhibit a more gradual and consistent increase in KS values, reflecting a smoother drift pattern.

The consistently low p-values ($<$ 0.05) further confirm that the detected distributional differences are unlikely to be due to random chance. For example, in the sixth month, the p-value for BloodMNIST drops to approximately 0.008, reinforcing the significance of the observed drift. These findings suggest that the datasets are diverging from their initial distributions in ways that are both measurable and meaningful.

The analysis of KS statistics and p-values from Fig. \ref{fig:multi_report} underscores the importance of robust drift detection mechanisms. While the KS test effectively highlights significant distributional changes, the observed fluctuations and consistently low p-values point to the limitations of raw feature extraction methods. These results emphasize the need for fine-tuning and additional techniques to better adapt to the specific characteristics of each dataset and ensure reliable drift detection.

 \begin{figure}[t]
    \centering
    \includegraphics[width=0.30\textwidth]{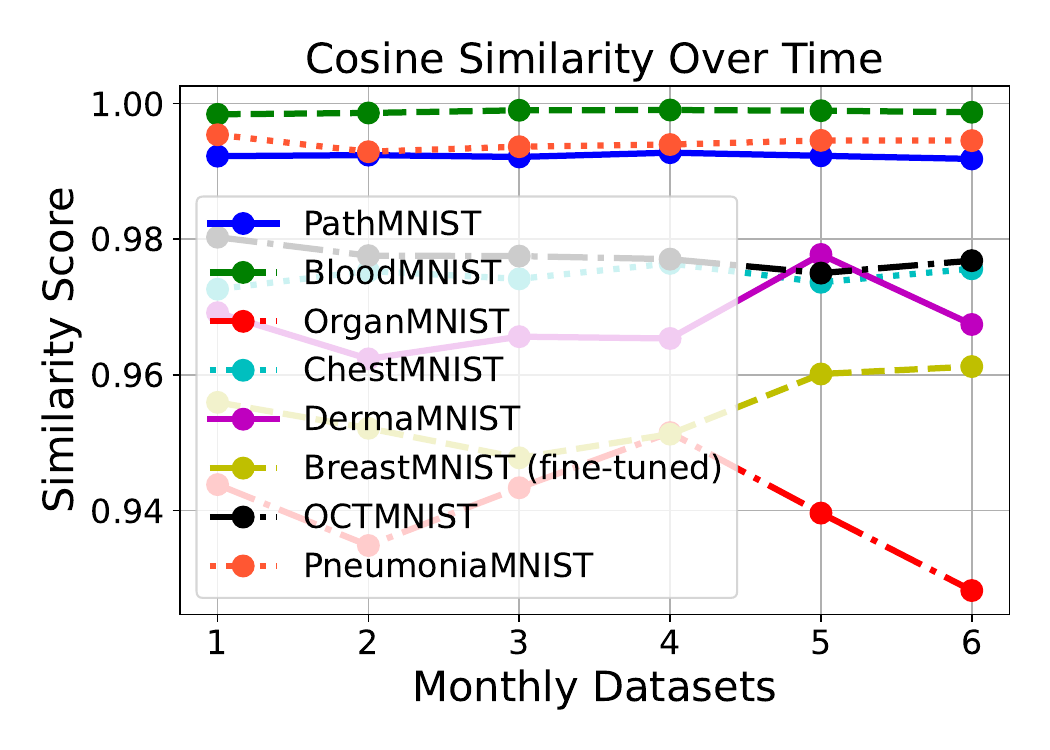}
    \caption{\textbf{Baseline 3 (fine-tuned without data-sketches and noises)}: Cosine similarity of fine-tuned model without data-sketches. We fine-tuned the BreastMNIST ViT model; however, without applying data sketches to remove noise, the similarity scores still fluctuate, making it difficult to reliably detect data drift.}
    \label{fig:report55556}
\end{figure}

\textbf{Baseline 2} (without data-sketches and fine-tuning). In Fig. \ref{fig:multi_report2}, it is clear that the cosine similarity scores across the 8 Med-MNIST datasets remain consistently low. This suggests that the feature extraction process, without fine-tuning, performs poorly in capturing meaningful similarities between highly similar images. Despite the datasets containing images that should be visually close, the cosine similarity scores indicate otherwise.

The poor similarity scores, around 45\% to 55\%, imply that the model, as it stands without fine-tuning, is not able to generate feature vectors that sufficiently differentiate between subtle variations in the image data. This is problematic, especially when dealing with medical imaging tasks where accurate feature representation is crucial for anomaly detection or diagnosis. The fact that even visually similar images yield low cosine similarity scores points to a fundamental issue in the feature extraction mechanism, underscoring the need for fine-tuning to improve the quality of these extracted features.

\textbf{Baseline 3} (fine-tuned without data-sketches and noises). In Fig. \ref{fig:report55556} illustrates that without data-sketches, there is noticeable fluctuation in the cosine similarity scores across different datasets over time. This fluctuation indicates inconsistencies in data quality, leading to uneven similarity comparisons. The variations suggest that certain datasets have more significant differences, resulting in a drop in similarity scores.

Incorporating data-sketches can help mitigate these fluctuations by making similarity comparisons more consistent and fair. Data-sketches allow for more compact and robust feature representations, reducing noise and outliers in the data. This ultimately leads to more reliable and stable similarity measurements across datasets, improving the overall fairness and accuracy of the comparisons.

\begin{figure*}[t]
    \centering
    \begin{subfigure}[b]{0.23\textwidth}
        \centering
        \includegraphics[width=\textwidth]{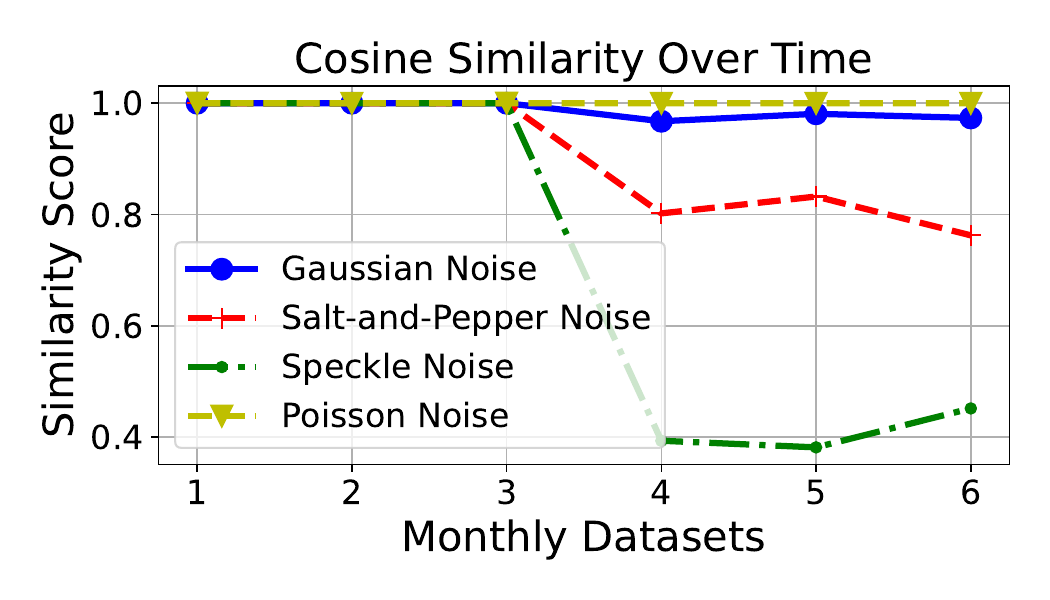}
        \caption{5\% Pixels Random Noise}
        \label{fig:report31}
    \end{subfigure}
    \hfill
    \begin{subfigure}[b]{0.23\textwidth}
        \centering
        \includegraphics[width=\textwidth]{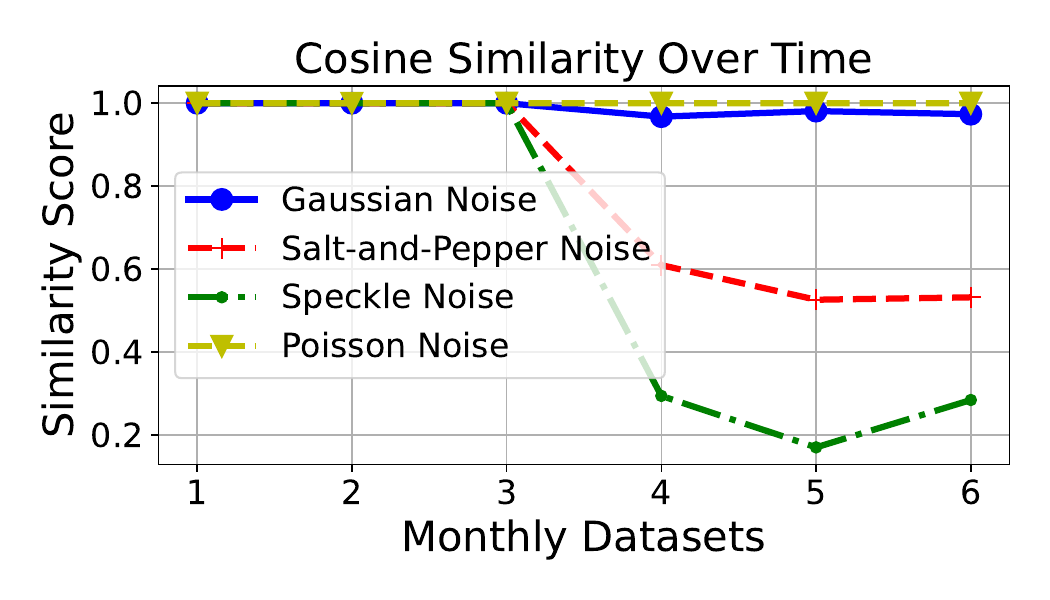}
        \caption{20\% Pixels Random Noise}
        \label{fig:report2}
    \end{subfigure}
    \hfill
    \begin{subfigure}[b]{0.23\textwidth}
        \centering
        \includegraphics[width=\textwidth]{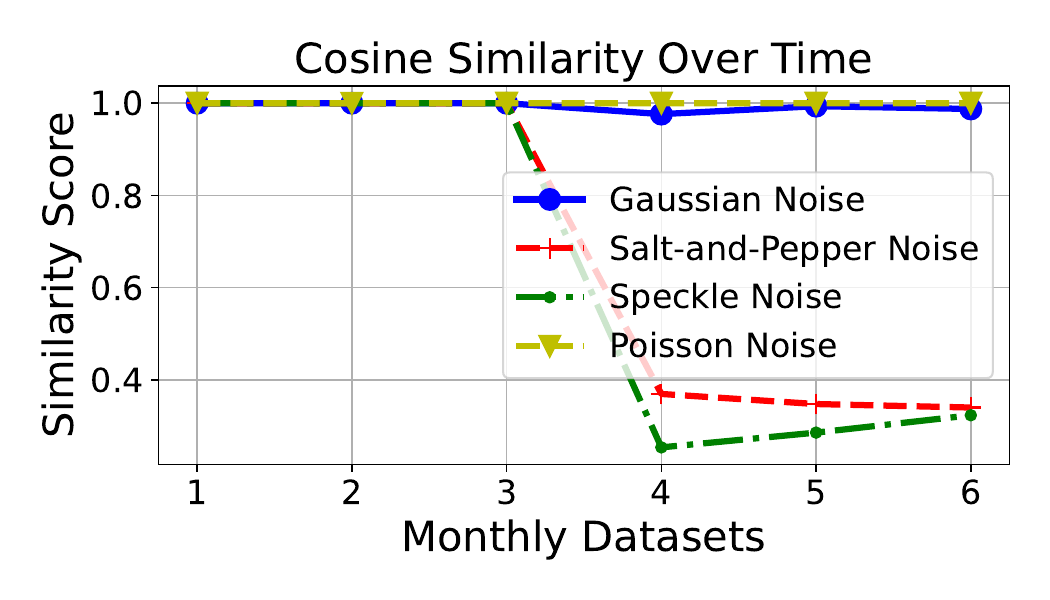}
        \caption{30\% Pixels Random Noise}
        \label{fig:report3}
    \end{subfigure}
    \hfill
    \begin{subfigure}[b]{0.23\textwidth}
        \centering
        \includegraphics[width=\textwidth]{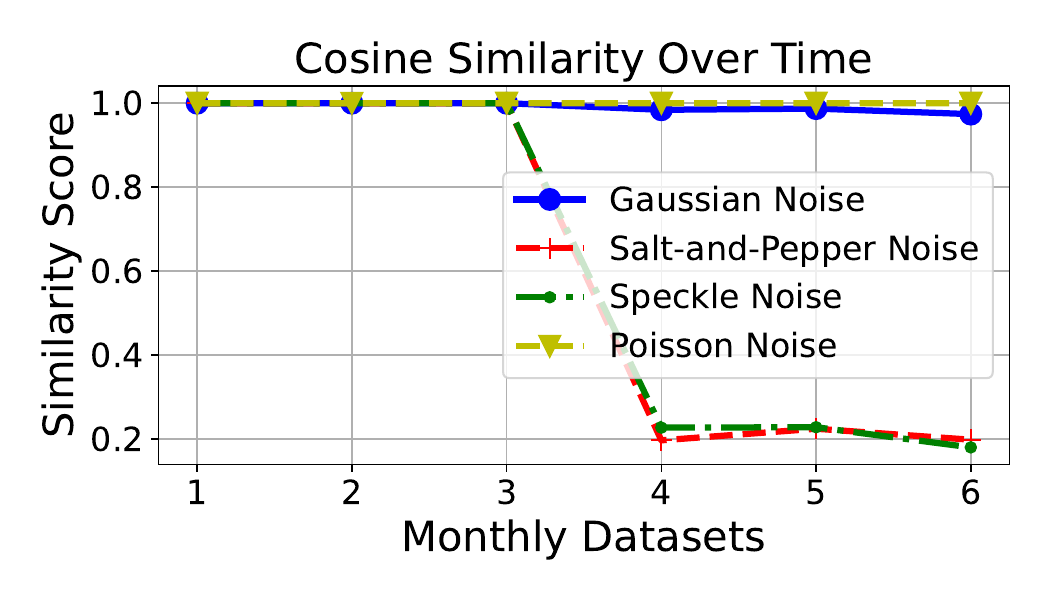}
        \caption{50\% Pixels Random Noise}
        \label{fig:report4}
    \end{subfigure}

    \begin{subfigure}[b]{0.23\textwidth}
        \centering
        \includegraphics[width=\textwidth]{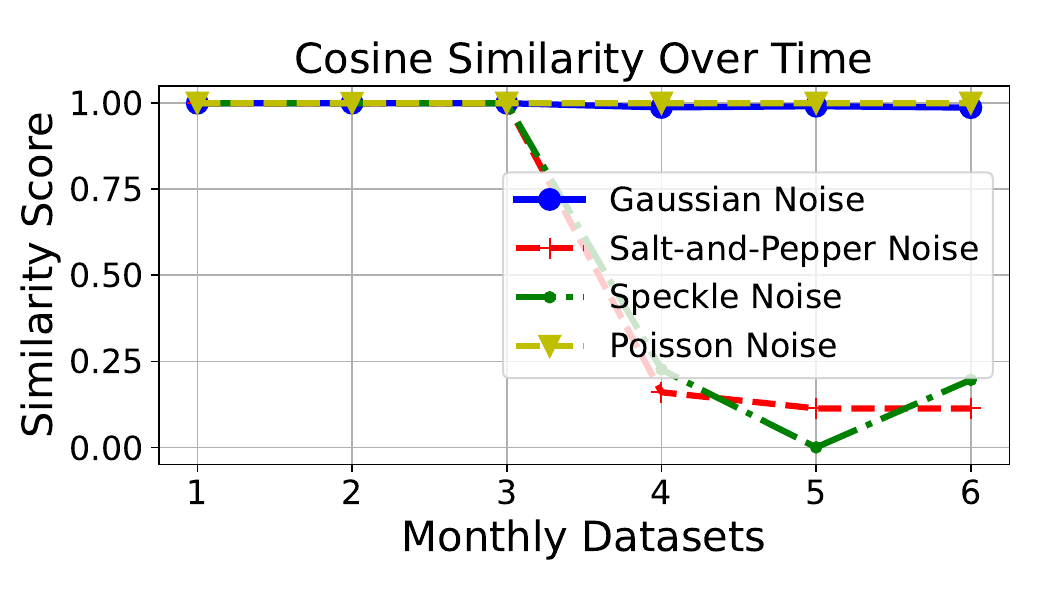}
        \caption{60\% Pixels Random Noise}
        \label{fig:report11123}
    \end{subfigure}
    \hfill
    \begin{subfigure}[b]{0.23\textwidth}
        \centering
        \includegraphics[width=\textwidth]{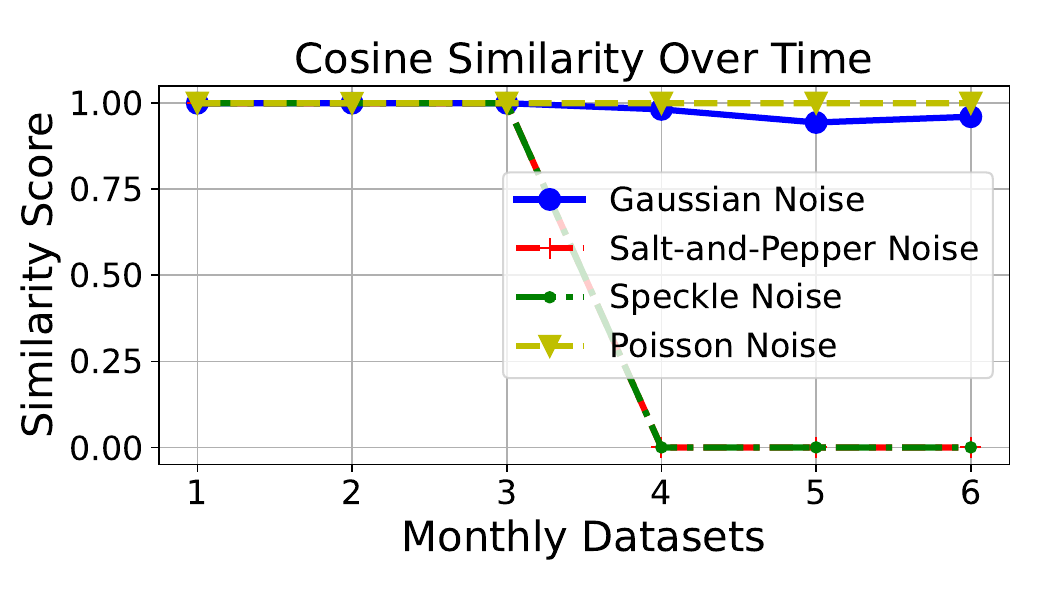}
        \caption{80\% Pixels Random Noise}
        \label{fig:report2}
    \end{subfigure}
    \hfill
    \begin{subfigure}[b]{0.23\textwidth}
        \centering
        \includegraphics[width=\textwidth]{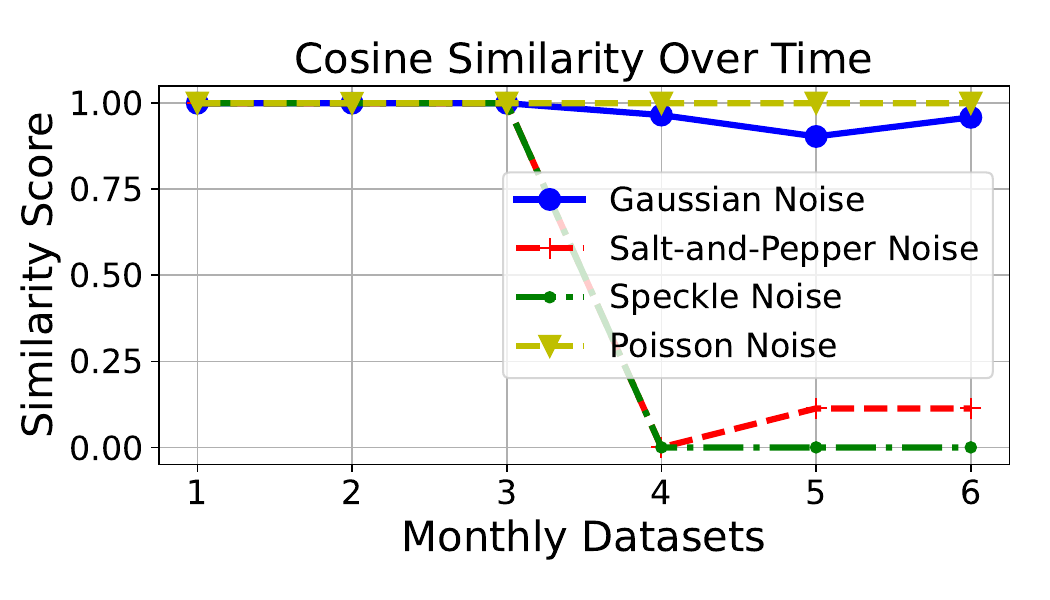}
        \caption{90\% Pixels Random Noise}
        \label{fig:report3}
    \end{subfigure}
    \hfill
    \begin{subfigure}[b]{0.23\textwidth}
        \centering
        \includegraphics[width=\textwidth]{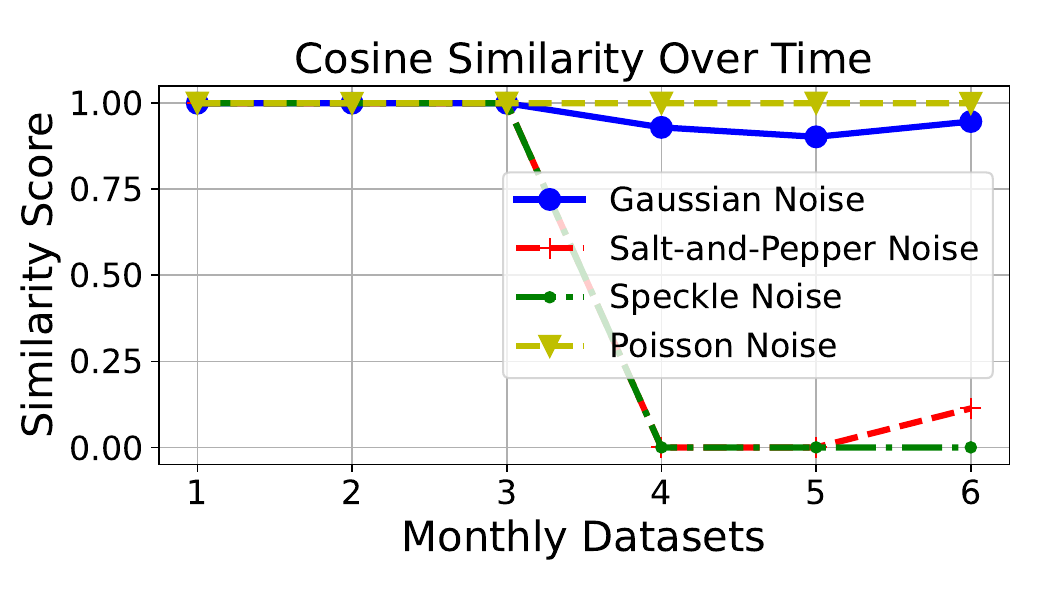}
        \caption{100\% Pixels Random Noise}
        \label{fig:report4}
    \end{subfigure}
    \caption{\textbf{Our Performance in Breast Pre-train Model with Data-sketches:} Cosine Similarity Score Across 8 Random Noises. The evaluation shows that our method maintains high cosine similarity scores (99\%) under low noise levels, demonstrating stability and robustness. As noise increases, the scores decline, but the method effectively detects shifts, showcasing its sensitivity to distributional changes.}
    \label{fig:multi_report3}
\end{figure*}

 \begin{figure}[t]
    \centering
    \includegraphics[width=0.27\textwidth]{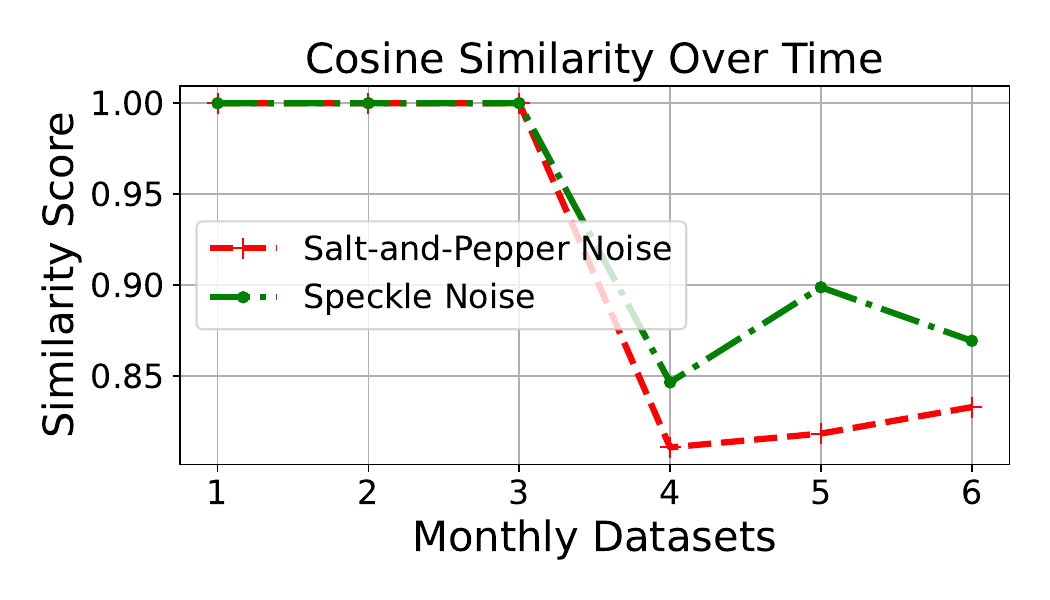}
    \caption{Drift detection in 1\% random noise.}
    \label{fig:report5555}
\end{figure}

\textbf{Our Performance} (fine-tuned with data-sketches). In Fig. \ref{fig:multi_report3}, the cosine similarity scores using BreastMNIST with varying levels of noise demonstrate that the feature extraction process is of high quality. For the first three monthly datasets, the similarity scores highly stable (\textbf{around 99\% similarity with no fluctuations}) across different datasets, indicating no significant drift and consistent feature representation. This stability reflects our solution's robustness in handling abnormal and bad quality datasets.

Starting from the fourth dataset, the similarity scores drop, signaling the onset of drift. Despite this, the model effectively detects these changes, showing its sensitivity to distributional shifts. The earlier high scores emphasize the model's strong performance in stable conditions, while its ability to identify drift highlights its suitability for real-time anomaly detection in medical imaging datasets.

\subsection{Sensitivity Evaluation}

In the background, Section \ref{factors}, we introduced several factors that can cause dataset drift. Here, we introduce four types of noise: Gaussian Noise, Salt-and-Pepper Noise, Speckle Noise, and Poisson Noise \cite{boncelet2009image}, each corresponding to one of these drift-inducing factors. These noises are used to test the sensitivity of our method under different conditions that may simulate real-world drift scenarios

\subsubsection{Gaussian Noise} Gaussian noise is often associated with sensor performance fluctuations or low-light conditions. It can occur due to electronic noise in the imaging sensor, which might be influenced by temperature or poor lighting.

\subsubsection{Salt-and-Pepper Noise}
This type of noise can be linked to \textit{transmission errors} or intentional tampering. It's a form of impulse noise where random pixels in the image are corrupted, leading to black-and-white spots, resembling salt and pepper.

\subsubsection{Speckle Noise}
Speckle noise is commonly observed in coherent imaging systems such as ultrasound, radar, or SAR. It results from the interference of wavefronts, leading to granular noise, and can be considered when evaluating \textit{subject variations} or \textit{device-specific characteristics}.

\subsubsection{Poisson Noise}
Poisson noise is related to the statistical nature of photon counting and is often linked to \textit{lighting conditions}. In low-light imaging scenarios, the number of photons hitting the sensor follows a Poisson distribution, leading to this type of noise. Therefore, it can be associated with varying lighting conditions or low-light environments.

Fig. \ref{fig:report5555} shows the cosine similarity scores over time with just 1\% random noise applied to the Med-MNIST datasets. The results highlight the sensitivity of the model, as even with this minimal noise level, the system is able to detect drift effectively. Notably, when drift occurs around the fourth dataset, the similarity score drops significantly from nearly 100\% to around 81\%.

This sharp decline demonstrates the model's capability to respond to even small distributional changes, making it highly sensitive to drift (with qualified datasets). The ability to detect such subtle shifts with only 1\% noise showcases the robustness of the approach, ensuring that even minor anomalies or deviations in the data can be identified promptly. This sensitivity is particularly valuable in medical imaging, where early detection of even the slightest anomalies can be critical for diagnosis and treatment.

\section{Conclusion} \label{con}
This paper presents an innovative approach to detecting distributional drift in medical imaging by combining data-sketching techniques with a fine-tuned Vision Transformer (ViT) model. Our method effectively addresses challenges such as high-dimensional image data, noisy datasets, and the need for real-time detection. By leveraging data sketches for compact, noise-resilient representations and fine-tuning ViT for enhanced feature extraction, the approach achieves exceptional sensitivity to subtle data shifts, as demonstrated by its 99.11\% accuracy on breast cancer CT image classification tasks. This work provides a robust and scalable solution for maintaining diagnostic model accuracy in dynamic clinical settings. Future research can explore extending this approach to other medical imaging modalities or integrating it with hospital workflows to improve practical applicability. By ensuring consistent and reliable performance, our method contributes to advancing AI-driven tools for better patient care and safety.

\section*{Acknowledgments}
This research was supported by the National Science Foundation under Award Number 2326034 as part of the NSF/FDA scholar-in-residence program, with valuable input from Regulatory Scientists at the FDA's DIDSR within CDRH’s OSEL. 

\bibliographystyle{ieeetr}
\bibliography{refs}

\end{document}